\documentclass[useAMS,usenatbib,usegraphicx]{mn2e}

\newcommand{\kev}{keV}
\newcommand{\ergcms}{erg~cm$^{-2}$~s$^{-1}$}
\newcommand{\fe}{Fe~K$\alpha$}
\newcommand{\etal}{et al.}
\newcommand{\fourc}{4C+74.26}
\newcommand{\osev}{O~\textsc{vii}}
\newcommand{\oeight}{O~\textsc{viii}}
\newcommand{\xmm}{\textit{XMM-Newton}}

\title[The \textit{XMM-Newton} view of \fourc]
  {A complete view of the broad-line radio galaxy \fourc\ with \xmm}
\author[D.\ R.\ Ballantyne]
  {D.~R.~Ballantyne \thanks{ballantyne@cita.utoronto.ca}\\
  $$Canadian Institute for Theoretical Astrophysics, 60 St. George
  Street, Toronto, Ontario, Canada M5S 3H8}

\pagerange{\pageref{firstpage}--\pageref{lastpage}}
\pubyear{2005}

\usepackage{times}

\begin{document}

\label{firstpage}

\maketitle

\begin{abstract}
This paper presents a timing study and broadband spectral analysis of
the broad-line radio galaxy \fourc\ based on a 35~ks \xmm\
observation. As found in previous datasets, the source exhibits no
evidence for rapid variability, and its 0.2--10~\kev\ lightcurve is
well fit by a constant. An excellent fit to the pn 0.3--12~\kev\
spectrum was found using a continuum that combines an ionized and a
neutral reflector, augmented by both cold and warm absorption. There
is no evidence for a soft excess. The column of cold absorption was
greater than the Galactic value with an intrinsic column of $\sim 1.9\times
10^{21}$~cm$^{-2}$. Evidence
for the warm absorber was found from \osev\ and \oeight\ absorption
edges with maximum optical depths of $\tau^{\mathrm{max}}_{\mathrm{O\
VII}}=0.3$ and $\tau^{\mathrm{max}}_{\mathrm{O\ VIII}}=0.03$,
respectively. A joint pn-MOS fit increased the \oeight\ optical depth
to $\tau^{\mathrm{max}}_{\mathrm{O\ VIII}}=0.1$. A simple, one-zone
warm absorber model yielded a column of $\sim 9\times
10^{20}$~cm$^{-2}$ and an ionization parameter of $\sim 60$. Partial
covering models provide significantly worse fits than ones including a
relativistically broadened \fe\ line, strengthening the case for the
existence of such a line. On the whole, the X-ray spectrum of \fourc\
exhibits many features typical of both a radio-loud quasar (excess
absorption) and radio-quiet Seyfert~1 galaxies (\fe\ emission and warm
absorption). We also show that a spurious absorption line at $\sim
8$~\kev\ can be created by the subtraction of an instrumental Cu
K$\alpha$ emission line.

\end{abstract}


\section{Introduction}
\label{sect:intro}
\fourc\ ($z=0.104$; \citealt{ril88}) is a luminous broad-line radio
galaxy (BLRG) most notable for its large radio lobes, extending 10
arcminutes (tip-to-tip) on the sky \citep{ril88}. A one-sided jet has
been observed with the VLA \citep{rw90} and on pc scales with VLBI
\citep{pea92}. The flux limit on a counterjet gives a limit to the
inclination angle of $i \la 49$~degrees \citep{pea92}, resulting in a
physical size for the radio source of $\sim 2$~Mpc (using a
$\textit{WMAP}$ cosmology: $H_0=70$~km~s$^{-1}$~Mpc$^{-1}$,
$\Omega_{\Lambda}=0.73$, $\Omega_0=1$; \citealt{spe03}), well within
the range to be classified as a giant radio galaxy
\citep{lar01,lar04}. The observed morphology of the radio jets clearly
places it as a FRII source, although the radio luminosity of \fourc\
is on the border between FRI and FRII \citep{ril88}. The bolometric
luminosity has been estimated to be $L_{\mathrm{bol}} \approx 2\times
10^{46}$~ergs~s$^{-1}$ \citep{wu02}, indicating that \fourc\ is also
close to the Seyfert-quasar border. The host galaxy has a size and
luminosity typical of other giant ellipticals associated with radio
sources \citep{ril88}. Optical spectra reveal very broad permitted
lines with measurements of the H$\beta$ FWHM ranging from $\sim
8000$~km~s$^{-1}$ \citep{ril88,brink98,rob99} to 11,000~km~s$^{-1}$
\citep{corb97}. Using this last value, \citet{wu02} employ the
broad-line region radius-luminosity relation \citep{kas00} to obtain a
black hole mass of $\sim 4\times 10^9$~M$_{\odot}$.

X-ray observations of \fourc\ began with a 23~ks \textit{ASCA}
observation in 1996. \citet{brink98} presented the first analysis of
these data, as well as \textit{ROSAT} data. The data from the
\textit{ROSAT} All-Sky Survey yielded a very hard photon-index for the
0.3--2~\kev\ band, $\Gamma=1.2-1.3$, with a cold absorption column
slightly above the Galactic value of $1.19\times 10^{21}$~cm$^{-2}$
\citep{dl90}. \textit{ROSAT} PSPC data of \fourc\ was available from a
20~ks observation of the cataclysmic variable VW Cep during which the
BLRG was in the field-of-view. An absorbed power-law fit to this
dataset by \citet{brink98} also revealed a hard power-law ($\Gamma
\sim 1.6$) with higher than Galactic absorption. A later analysis of
this PSPC data by \citet{km00} found that a dusty warm absorber model
provided a good fit and increased the photon-index to values closer to
those found in the \textit{ASCA} data ($\Gamma \approx 2$) and in
\textit{ROSAT} samples of radio-loud quasars \citep[e.g.][]{bys97}.

The \textit{ASCA} data has subsequently been re-analyzed by
\citet*{sem99} and \citet{rt00}. All groups found a best fit model
that included a power-law with a photon index $\Gamma \approx 2$ and
cold absorption in excess of the Galactic value. However,
\citet{brink98} preferred a solution with a warm absorber, while the
best fit of \citet{sem99} did not require one ($\tau_{\mathrm{O\ VII}}
< 0.8$). Similarly, \citet{brink98} and \citet{rt00} find a Gaussian
\fe\ line at the 97 per cent confidence level, but \citet{sem99} find
one about twice as strong and at $>99$ per cent confidence. The
\textit{ASCA} data also show a hardening at high energies, which
\citet{brink98} and \citet{rt00} model with a reflection continuum,
but find very high reflection fractions ($R \sim 6$ and $3$,
respectively). \citet{sem99} argue that such a large reflection
fraction should yield a much stronger \fe\ line, and propose a model
where the hardening is due to a second, very hard power-law with
$\Gamma \sim 0.2$, possibly arising from the radio jet.

This confusing situation was improved by a 100~ks \textit{BeppoSAX}
observation presented by \citet*{hse02}. No significant variability
was detected from \fourc\ even in the high-energy PDS band. This fact
provides strong evidence that jet emission is not significantly
contributing to the hard X-rays. Indeed, \citet{hse02} find that
Compton reflection provides the best fit to their broadband data with
$R \sim 1$. These authors also find cold absorption in excess of the
Galactic value, but a warm absorber is not required by the data. A
significant, but unresolved, \fe\ line was also detected in the
\textit{BeppoSAX} spectrum.

More recently, very preliminary results from a 70~ks \textit{Chandra}
gratings observation of \fourc\ have been presented by
\citet{kas04}. A highly ionized and weak warm absorber was detected by
\textit{Chandra}, including emission and absorption lines from H-like
and He-like Mg, Al and Si. Moreover, the low number of counts at
energies less than 0.8~\kev\ could indicate \osev\ and \oeight\ edges
(Kaspi, private communication).

Here, we present the results of a 35~ks \xmm\ observation of \fourc\
which will for the first time properly characterize the X-ray spectrum
of this source. The detection of a relativistically broadened \fe\
line has already been reported elsewhere (\citealt{bf05}; hereafter
Paper I). This paper therefore concentrates on the timing analysis
(Sect.~\ref{sect:timing}) and fitting the broadband spectrum
(Sect.~\ref{sect:spectral}). The paper concludes by discussing the
results in Sect.~\ref{sect:discuss}. We begin in the next section by
describing the details of the observation and data reduction.

\section{Observations and Data Reduction}
\label{sect:obs} 
\fourc\ was observed by \xmm\ \citep{jan01} for 35~ks during
revolution 762 starting at 2004 February 6 13:57:42. Data were
collected using the single pn \citep{str01} and two MOS \citep{tur01}
detectors in the European Photon Imaging Camera (EPIC) system, both
Reflection Grating Spectrometers (RGS; \citealt{dh01}) and the Optical
Monitor (OM; \citealt{mas01}). The EPIC instruments were operated in
large-window mode with the medium optical filter in place. The RGS was
operated in standard spectroscopy mode.

Data reduction was performed using the \xmm\ Science Analysis System
(SAS) v.6.1. The analysis chains \textsc{epchain} and \textsc{emchain}
were run on the observation data files to produce calibrated event
lists for the MOS and pn detectors by removing bad pixels and applying
both gain and Charge Transfer Inefficiency (CTI) corrections to the
data\footnote{see, e.g.,
http://xmm.vilspa.esa.es/sas/current/doc/epchain/index.html}. Source
spectra were extracted using circles of radius 115 arcseconds (for the
pn), 119 arcseconds (MOS1) and 122 arcseconds (MOS2). Background
spectra were extracted from source free areas on the same CCD using
circles with radii of 60 arcseconds (pn and MOS1) and 50 arcseconds
(MOS2). The extracted pn spectrum included both single and double
events, while the MOS spectra were comprised of events with all
patterns.  The background was negligible during the observation except
for a minor enhancement at $\sim 22$~ks into the integration. A
good-time interval file was constructed as described in \S~4.4.3 of
the \xmm\ SAS User's
Guide\footnote{http://xmm.vilspa.esa.es/external/xmm\_user\_support/documentation/\\\phantom{xxhttp://}sas\_usg\_3.0/USG/USG.html}
to remove any potential contamination by background events. The SAS
task \textsc{epatplot} was used to check the pattern distributions in
the EPIC data, and both MOS spectra were found to suffer from a
non-negligible amount of pileup \citep{bal99}. To correct this, the
MOS spectra were re-extracted using only single (i.e., pattern 0)
events. Following background subtraction, the final pn spectrum has
28.8~ks of good exposure time and consists of over $2.5\times 10^5$
counts, giving a mean count rate of 8.6~s$^{-1}$. The final background
subtracted MOS-1 and MOS-2 spectra each had mean count rates of
2.2~s$^{-1}$, and contained $7.4\times 10^4$ and $7.5\times 10^4$
counts, respectively. Prior to spectral analysis, all data were
grouped using \textsc{grppha} to have a minimum of 20 counts per
bin. Finally, the SAS tasks \textsc{rmfgen} and \textsc{arfgen} were
utilized to produce the response matrix and ancillary response files.

We note that the background extraction regions used above are smaller
than the source extraction regions. This was done in order that the
background was taken from either the same CCD or same window as the
majority of the source counts. To check the spectral analysis described in the
following section, larger background regions from a different CCD with
radii equal to or larger than the source regions were also
extracted. The derived spectral parameters were not significantly
changed by using the larger background regions, therefore we report
the results from the original data extraction.

The RGS data were reduced using the \textsc{rgsproc} chain in the
SAS. The observed count rate was 0.15~s$^{-1}$ for RGS1 and
0.19~s$^{-1}$ for RGS2. This yielded only $5.6\times 10^3$ and $6.8\times
10^3$ counts in the RGS1 and RGS2 spectra, respectively. Given the
small number of counts it was decided not to pursue a detailed
analysis of the RGS spectra, although the continua predicted by the
broadband pn models (including the warm absorption edges) were checked
against the observed RGS spectral shape. 

\section{Timing analysis}
\label{sect:timing}
\begin{figure}
\centerline{
\includegraphics[width=0.35\textwidth,angle=-90]{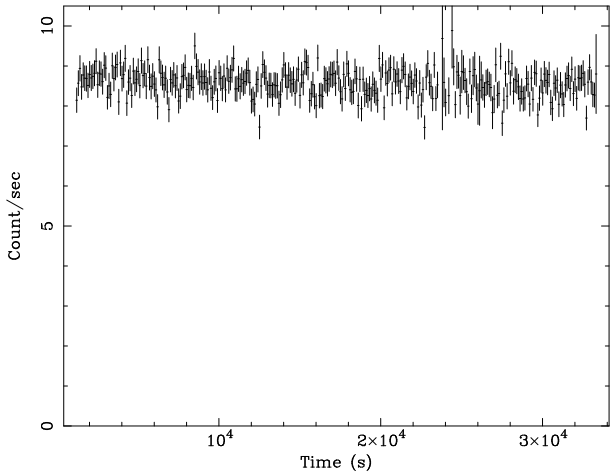}
}
\caption{The 0.2--10~\kev\ pn lightcurve of \fourc\ plotted using
  100~s time bins. The data were background subtracted and corrected
  for vignetting and deadtime losses.}
\label{fig:lightcurve}
\end{figure}

Figure~\ref{fig:lightcurve} shows the 0.2--10~\kev\ pn lightcurve of
\fourc\ in 100~s time bins. The SAS task \textsc{lccorr} was used to
background subtract the data, as well as provide corrections for
vignetting and deadtime.
As with the earlier \textit{ASCA} \citep{brink98} and
\textit{BeppoSAX} \citep{hse02} observations, no significant
variability was observed from \fourc. A constant fit to the pn
lightcurve resulted in a $\chi^2/$d.o.f.$=267/317$ (d.o.f.=degrees of
freedom), with a best fit of 8.57~s$^{-1}$.

\section{Spectral Analysis}
\label{sect:spectral}
In this section we present the results of fitting the \xmm\ spectrum
of \fourc\ between 0.3 and 12~\kev\ in the observed frame. Since it
has the largest number of counts and higher spectral resolution, the
pn spectrum was initially analyzed on its own, but we include the MOS
data (from 0.3--10~\kev) at the end of this section to check for any
differences. As mentioned above, results for the \fe\ line region and
reflection parameters are presented in Paper I. Here, we will
concentrate on the absorption characteristics in the observed
spectrum. XSPEC v.11.3.1p \citep{arn96} was used for the spectral
fitting, and a $\Delta \chi^2=2.71$ criterion was used to determine
the 2$\sigma$ errorbars on the best-fit parameters. Galactic
absorption, modeled using the 'TBabs' code in XSPEC \citep{wam00}, is
included in all fits. Unless stated otherwise, the figures are plotted
in the observed frame, while the fit parameters are quoted in the rest
frame.

\begin{figure}
\centerline{
\includegraphics[width=0.32\textwidth,angle=-90]{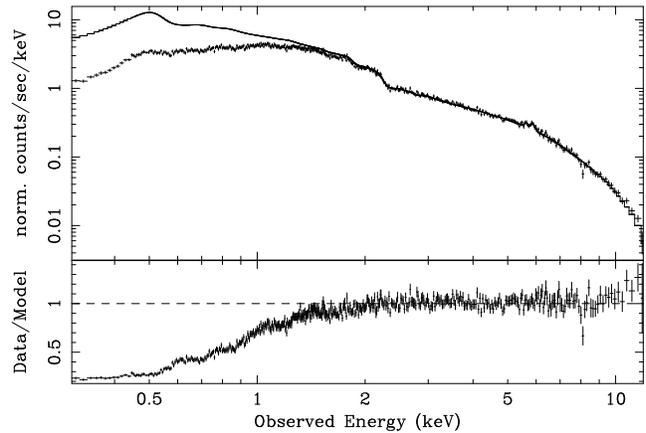}
}
\caption{The top panel shows the pn count spectrum of \fourc, while
  the lower panel displays the data-to-model ratio between 0.3 and
  12~\kev\ when the model (solid line in the top panel)
  IONDISK*blr+IONDISK+G from Paper I is extended to low
  energies. Significant absorption is required at energies below
  2~\kev.}
\label{fig:ratio}
\end{figure}

We begin the analysis with the best-fitting model found in fitting the
2--12~\kev\ data. This model, denoted IONDISK*blr+IONDISK+G in Paper
I, consists of a relativistically blurred ionized disk (employing the
models of \citealt*{rfy99}), an unblurred neutral reflector (also
using the \citealt{rfy99} models), and a narrow Gaussian emission
line. The data-to-model ratio when the 0.3--2~\kev\ data are included
is shown in Figure~\ref{fig:ratio}.
The plot clearly shows that a significant amount of absorption is
required at energies $< 2$~\kev. This opacity could be in the form of
extra cold absorption (as favored by the earlier analyses of
\citealt{sem99} and \citealt{hse02}), warm absorption (as indicated by
\citealt{kas04}), or a combination of the two.

\begin{figure}
\centerline{
\includegraphics[width=0.32\textwidth,angle=-90]{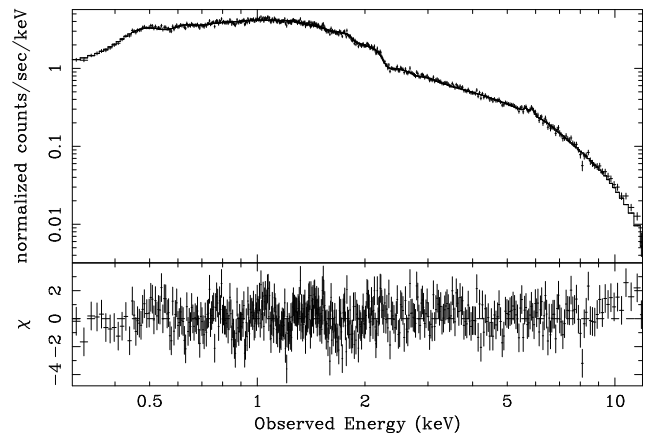}
}
\caption{The top panel shows the pn count spectrum of \fourc\ with the
    solid line denoting our best fit model (2nd line in
    Table~\ref{table:fit1}). The residuals to the fit in units of
    standard deviations are shown in the lower panel. There is no
    evidence for a soft excess.}
\label{fig:ratio2}
\end{figure}

\begin{figure}
\centerline{
\includegraphics[width=0.32\textwidth,angle=-90]{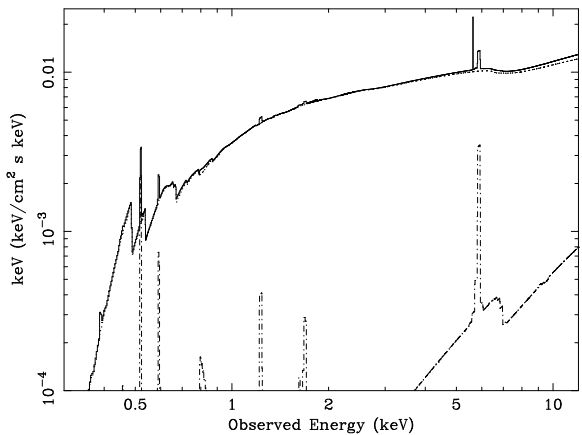}
}
\caption{The best fitting model to the
  0.3--12~\kev\ spectrum of \fourc\ (solid line; second line in
  Table~\ref{table:fit1}). The continuum is comprised of a
    relativistically blurred ionized reflector (dot-dashed line) and a
    neutral reflector (dashed line).}
\label{fig:model}
\end{figure}

To test the different absorption models, we first froze the
parameters in the model related to the \fe\ lines, and then added
additional intrinsic absorption via the 'ztbabs' model in XSPEC. A
good fit was found ($\chi^2/$d.o.f.=1643/1643) to the 0.3--12~\kev\
spectrum with an intrinsic column
$N_{\mathrm{H}}^{\mathrm{intr}}=1.83\times 10^{21}$~cm$^{-2}$. When
the cold absorption model was replaced with the warm absorber model
'absori' \citep{don92}, the ionization parameter of the absorber went to zero
and basically the same fit was uncovered as before
($\chi^2/$d.o.f.=1638/1642). However, the \textit{Chandra} data of
\fourc\ show evidence for a weak warm absorber with possible ionized
oxygen edges. To check this, an \oeight\ edge at 0.87~\kev\ and
an \osev\ edge at 0.739~\kev\ were added to the cold absorption model
above. This new model resulted in a significant ($\Delta
\chi^2=-54$ with only 2 additional degrees of freedom) improvement
over the purely cold absorber case, bolstering the case for a warm
absorber toward \fourc. Since the previous model excluding the
warm absorber resulted in a statistically acceptable fit, it is not
possible to conclude that the data require warm absorption. The
results of the fit are shown in the first line of Table~\ref{table:fit1}.
\begin{table*}
\begin{minipage}{152mm}
\caption{Results from fitting the spectrum of \fourc\ from 0.3 to
  12~\kev. The spectral model includes two absorption edges fixed at
  the energies of \osev\ and \oeight\ (0.739~\kev\ and 0.87~\kev,
  respectively), intrinsic cold absorption, Galactic absorption, and
  the double reflector plus Gaussian model from Paper I. In the first
  line of the table, the parameters affecting the \fe\ line were
  frozen at their best fit values from Paper I. The final row presents
  the results from a joint pn-MOS fit, where the MOS data were used
  between 0.3 and 10~\kev. In this case, only the photon-index from
  the pn data is reported. $\Gamma$ is the
  photon-index of the incident power-law,
  $N_{\mathrm{H}}^{\mathrm{intr}}$ is the column of the intrinsic cold
  absorber (cm$^{-2}$), $\xi$ is the ionization parameter of the
  ionized reflector which makes up most of the continuum,
  $\tau^{\mathrm{max}}_{\mathrm{O\ VIII}}$ and
  $\tau^{\mathrm{max}}_{\mathrm{O\ VII}}$ are the maximum optical
  depths in the \oeight\ and \osev\ edges respectively,
  $r_{\mathrm{out}}$ is the outer radius of the blurred ionized
  reflector (gravitational radii), $i$ is the inclination angle
  (degrees), and $E$ is the energy of the narrow Gaussian emission
  line (keV).}
\label{table:fit1}
\begin{tabular}{ccccccccc}
$\Gamma$ & $N_{\mathrm{H}}^{\mathrm{intr}}$ & $\log \xi$ &
    $\tau^{\mathrm{max}}_{\mathrm{O\ VIII}}$ &
    $\tau^{\mathrm{max}}_{\mathrm{O\ VII}}$ & $r_{\mathrm{out}}$ & $i$
    & $E$ & $\chi^2/$d.o.f. \\
    \hline
1.86$\pm 0.01$ & $(1.87\pm 0.05)\times 10^{21}$ & 2.60$^{+0.06}_{-0.04}$
    & 0.03$^{+0.04}_{-0.03}$ & 0.17$^{+0.02}_{-0.03}$ & 6.3$^f$ &
    34$^f$ & 6.23$^f$ & 1589/1641 \\
1.85$\pm 0.01$ & $(1.89\pm 0.05)\times 10^{21}$ &
    2.63$^{+0.05}_{-0.06}$ & 0.03$\pm 0.03$ & 0.29$^{+0.04}_{-0.03}$ &
    6.3$^{+4.3}_{-2.6}$ & 43$^{+16}_{-3}$ & 6.22$^{+0.05}_{-0.04}$ &
    1542/1637 \\
1.89$\pm 0.01$ & $(1.98^{+0.5}_{-0.6})\times 10^{21}$ &
    2.71$^{+0.05}_{-0.02}$ & 0.10$^{+0.03}_{-0.02}$ &
    0.32$^{+0.03}_{-0.04}$ & 6.2 & 44$^{+15}_{-2}$ & 6.23$\pm 0.04$ &
    2503/2531 \\
\end{tabular}
\textit{Notes}: $^f$ parameter fixed at value
\end{minipage}
\end{table*}
There is strong evidence for an \osev\ edge in the data with a maximum
optical depth of $\sim 0.2$--$0.3$, consistent with the limit from the
\textit{ASCA} data. A further improvement to the fit was made when the
\fe\ line parameters were allowed to vary (second line in
Table~\ref{table:fit1}). Except for $\tau^{\mathrm{max}}_{\mathrm{O\
    VII}}$, the absorption parameters did not change
significantly. The residuals to this last fit (our best-fit model) are
shown in Figure~\ref{fig:ratio2}, while the model is displayed in Figure~\ref{fig:model}. 
The residuals show no evidence for a soft excess. The neutral
reflector predicts soft X-ray emission lines from \osev\ and \oeight\
that do seem to be accounted for in the data. It would be interesting
to obtain a long gratings observation with either \xmm\ or
\textit{Chandra} to confirm this possibility.

The above fits give strong indications that a warm absorber is present
in \fourc, but unfortunately cannot give much further information. In
order to try to further characterize the warm absorber, we performed a
fit where we replaced the absorption edges with the 'absori' model
(the extra cold absorption was still included in the model). To
simplify the procedure, the fit was performed with the \fe\ line
parameters frozen as in the first line of Table~\ref{table:fit1}. The
fit assumed solar abundances and an absorber temperature of $10^4$~K
(higher temperatures did not improve the results). A good fit was
obtained ($\chi^2/$d.o.f.=1616/1641) with a warm absorber column of
$8.8^{+2.4}_{-3.3}\times 10^{20}$~cm$^{-2}$ and an ionization
parameter of 56$^{+70}_{-44}$. The intrinsic cold absorption column
remained at $\sim 1.9\times 10^{21}$~cm$^{-2}$. While this
parameterization is clearly inferior compared to a gratings analysis,
it is consistent with Kaspi's early inferences of the warm absorber in
\fourc.

Using the best-fit model from Table~\ref{table:fit1}, we find a
0.5--2~\kev\ flux of $8.44\times
10^{-12}$~\ergcms\ and a 0.5--10~\kev\ flux of $3.26\times
10^{-11}$~\ergcms. The total unabsorbed luminosity between 0.5 and
2 (10)~\kev\ is $5.81\times 10^{44}$ ($1.25\times
10^{45}$)~ergs~s$^{-1}$. Employing the black hole mass
estimate of $\sim 4\times 10^9$~M$_{\odot}$ \citep{wu02}, and assuming
that the X-ray luminosity is $\sim 10$ per cent of the bolometric
luminosity, then \fourc\ has an Eddington ratio of $\sim 0.04$.

As a final check on the results, MOS data between 0.3 and 10~\kev\
were added and joint pn-MOS spectral fits were performed. The
normalizations of the MOS spectra were allowed to float relative to the
pn spectrum to account for any slight calibration errors. Also, a
separate photon-index was fit to the MOS data, as it is known that MOS
spectra are slightly harder than ones observed from the pn
\citep{vf04}. The joint pn-MOS fits resulted in a slightly steeper
$\Gamma$ than previously found, and consequently more absorption in
the edges and intrinsic cold absorber was required. The parameter
values from our best fit model are shown in the last line in
Table~\ref{table:fit1}. One can see there is only minor changes in the
fit parameters. When the 'absori' model was used, the warm absorber
column reduced to $7\times 10^{20}$~cm$^{-2}$, and the ionization
parameter also fell to $45$. Both these values are well within the
errorbars of the previous ones found using only the pn data.

\subsection{Partial covering models}
\label{sub:partialcovering}
Paper I presented evidence for a relativistically broadened \fe\ in
the 2--12~\kev\ spectrum of \fourc. The broad line was the
statistically preferred fit, but as seen in other AGN
\citep{sch03,ree04,gal04,tur05} the breadth of the \fe\ may also be
explained by a complicated absorption model. In this section we test
if complex absorption models can provide a better fit to the broadband
data than one involving a broad \fe\ line. 

The pn data are employed between 0.3--12~\kev\ to make use of its
superior sensitivity at 6~\kev. In contrast to NGC~3783
\citep{ree04} and NGC~3516 \citep{tur05}, \fourc\ does not have a very
significant warm absorber. The highly ionized component detected by
\textit{Chandra} was termed as ``weak'' (Kaspi, private
communication). Furthermore, the column estimated from the 'absori'
model is one to two orders of magnitude lower than the columns found
in NGC~3783 or NGC~3516. Therefore, it seems unlikely that curvature caused
by ionized absorption will significantly effect the spectrum at
energies close to the iron line. As a result, we concentrate on partial
covering models \citep{hol80} and use the 'zpcfabs' model within XSPEC
to simulate the effect of the primary X-ray continuum passing through
a neutral absorber with hydrogen column density $N_{\mathrm{pc}}$ that
covers a fraction $f_{\mathrm{pc}}$ of the X-ray source. The
attenuated continuum is assumed to consist of a reflection spectrum,
modeled with the 'pexrav' code of \citet{mz95} (same parameter values
as in Paper I), and an intrinsically
narrow ($\sigma=0.01$~keV) neutral \fe\ line. The warm absorber is
modeled with the \osev\ and \oeight\ edges, and Galactic absorption
was also included. The subsequent fit was poor with
$\chi^2/$d.o.f.=1752/1640, $N_{\mathrm{pc}} \approx 8\times
10^{20}$~cm$^{-2}$, and $f_{\mathrm{pc}}=0.95$. An improved fit was
found by adding additional cold absorption with the 'ztbabs' model. In
this case, $\chi^2/$d.o.f.=1674/1639, $N_{\mathrm{pc}} \approx 4\times
10^{22}$~cm$^{-2}$ and $f_{\mathrm{pc}}=0.25$. This fit also gave a
photon-index of $\Gamma=2.1$ and a reflection fraction of 2.5. The
column of extra cold absorption was $1.2\times 10^{21}$~cm$^{-2}$. A
further improvement to $\chi^2/$d.o.f.=1631/1640 was found by
replacing the 'pexrav' and Gaussian models with a solar abundance \citet{rfy99}
reflector that had its ionization parameter frozen at its lowest
value. In this case, $N_{\mathrm{pc}} \approx 7\times
10^{22}$~cm$^{-2}$ and $f_{\mathrm{pc}}=0.22$. The photon-index was
$\Gamma=1.93$ and a low reflection fraction of $0.4$ was also found. 

In summary, all the partial covering models considered here provide
poorer fits to the 0.3--12~\kev\ pn spectrum of \fourc\ than the ones
that included a relativistically broadened \fe\ line. The line should
still be confirmed with a higher signal-to-noise observation, but
these results help strengthen the conclusions presented in Paper I.

\subsection{A spurious high-energy absorption line}
\label{sub:absline}
Figures~\ref{fig:ratio} and~\ref{fig:ratio2} both show a possible
absorption line in the residuals at $\sim 8$~\kev. A narrow
($\sigma=0$) Gaussian absorption line added to the best-fit model
described above resulted in a $\Delta \chi^2=-10$ with 2 extra degrees
of freedom, significant at the $99.5$~per cent level according to the
F-test. However, this line is not of an astrophysical origin. It
arises from the subtraction of a strong Cu~K$\alpha$ emission line in
the background spectrum produced by the circuit board supporting the
pn detector in the spacecraft (\citealt{kat04} and references therein). It is important to note that the
strength of the this line is not constant over the instrument. In
fact, it is practically absent near the center where the image of the
target source falls. But, if the background accumulation region is
closer to the edge of a CCD (as it was in this case), it will include
this line which will then produce a spurious absorption feature in the
source data. Interestingly, if the line was interpreted as
Fe~\textsc{xxvi}~Ly$\alpha$, whose rest energy is 6.97~\kev, then one
would conclude that it was blueshifted to a velocity of $\sim
0.2c$. Thus, this feature could easily be misidentified as a highly
ionized and rapidly outflowing absorption feature.

As the Cu K$\alpha$ line is very narrow, when our \fourc\ spectral analysis
was repeated with a different background region that omitted the line,
the results did not significantly change. We therefore included the
affected plots in this paper to illustrate the potential danger to
future authors.

\section{Discussion}
\label{sect:discuss}
This paper presented the first detailed characterization of the
broadband X-ray spectrum of \fourc. At energies less than 2~\kev, the
spectrum is dominated by cold absorption in excess of the Galactic
column in this direction. If this extra attenuation is at the redshift
of \fourc, then the column required is $\sim 1.9\times
10^{21}$~cm$^{-2}$. This value is about two times smaller than the
previous estimates based on either \textit{ASCA} or \textit{BeppoSAX}
data \citep{sem99,hse02}, although it is not too different from the
\citet{rt00} value. Differing spectral models, analysis techniques and
the superior quality of the \xmm\ data are most likely responsible for
these disagreements. An excess column of cold absorption appears to be
common in radio-loud AGN \citep{sem99,rsg05}, and \fourc\ is no
different despite exhibiting typical Seyfert properties in almost
every other regard. In this case, perhaps the larger inclination
angle of $35$--$40$~degrees (as inferred from the \fe\ emission line;
Paper I \& Table~\ref{table:fit1}) implies that the line-of-sight to
the central engine has a longer path length within the host
galaxy. Alternatively, the host galaxy of \fourc, like most radio-loud
AGN, is an elliptical; therefore, unlike spiral galaxies, random
lines-of-sights into the galaxy are unlikely to have a clear path to
the center.

In addition to the excess cold absorption, \fourc\ also exhibits a
weak warm absorber, as primarily evidenced by a significant \osev\
edge. While a proper parametrization of the warm absorber awaits a
long gratings observation, we estimated the absorbing column and
ionization parameter to be $\sim 9\times 10^{20}$~cm$^{-2}$ and $\sim
60$, respectively. The high-throughput capabilities of \xmm\ are
allowing the discovery of warm absorbers in many more quasars than
before \citep{pic05}, emphasizing that they are a common occurrence in
all accreting supermassive black holes.

This analysis of the soft X-ray spectrum of \fourc\ does offer one
surprise that will need to be followed up with a longer
observation. The best fit broadband spectral model shown in
Fig.~\ref{fig:model} includes a very weakly ionized reflector which
predicts a number of recombination lines a low energies
\citep[e.g.][]{rfy99}, in particular from \osev\ and \oeight. These
lines do seem to fit the data, but whether they originate in the
accretion disk or warm absorber (or both) is unknown. This may be
elucidated with a long gratings observation of \fourc\ which would
unravel the properties of the warm absorber.

\section*{Acknowledgments}
Based on observations obtained with \textit{XMM-Newton}, an ESA
science mission with instruments and contributions directly funded by
ESA Member States and the USA (NASA). The author acknowledges
financial support from the Natural Sciences and Engineering Research
Council of Canada, and thanks J. Golding for help with the data
processing and analysis.


\bsp 

\label{lastpage}

\end{document}